\documentclass{aa}
\usepackage{graphicx}

\begin{document}

\title{Luminosity Function of GRBs}
\author{Shiv Sethi
        \inst{1}
        \and
        S. G. Bhargavi\inst{2}
       }

\institute{Harish-chandra Research Institute, Jhusi Road,
           Allahabad 211 019 India \\
           \email{sethi@mri.ernet.in}
           \and
          Indian Institute of Astrophysics, Sarjapur Road,
          Bangalore 560 034 India\\
           \email{bhargavi@iiap.ernet.in}
          }
\date{Received:        ; accepted:    }

\abstract{
We attempt to constrain the luminosity function of 
Gamma Ray Bursts (GRBs) from the observed number count--flux
relation and the afterglow redshift data. We assume three classes of 
luminosity functions for our analysis: (a) Log-normal distribution,
(b) Schechter distribution, and (c) Scale-free distribution.
We assume several models of the evolution of the GRB population 
for each luminosity function. 
Our analysis shows that: (a) log-normal is the only luminosity function
that is compatible with both the observations. This result is independent
of the GRB evolution model, (b) for log-normal function, the  
average photon luminosity $L_0$ and the width of the luminosity function
$\sigma$ that are compatible with both the observations fall in the 
range:  $10^{55} \, {\rm sec^{-1}} \la L_0 \la  10^{56} \, {\rm sec^{-1}}$ and $
2 \la \sigma \la 3$, (c) the agreement of observations with other
luminosity functions requires the GRB population to evolve more strongly
than the evolution of the star-formation rate of the universe. 
 }

\maketitle

\section{Introduction}

Recent afterglow observations of GRBs have given important clues about
the cosmological nature of GRBs and  their environment and  luminosities
(for details see Kulkarni {\it et al.} 2000 and
references therein).  
The afterglow database (see Greiner's
home page; Greiner (2000)) is sufficiently large that one could now think of
using these data for other studies in astronomy. For instance, one could
address the issue concerning the luminosity distribution of GRBs using
the redshift measurements available in the database.
The luminosity distribution of GRBs was hitherto obtainable only from
the number count-flux relationship (Piran 1992, 1999; Mao \& Paczy\'nski 1992;
Woods \& Loeb 1995, Lubin \& Wijers 1993; Ulmer \& Wijers 1995; Ulmer {\it et al.} 1995;
Cohen \& Piran 1995, Band,  Hartman,  \&  Schaefer 1999, Hakkila {\it
  et al.} 1996, Horack \& Hakkila 1997).  The redshift measurement of several GRB afterglows
  has allowed direct determination of GRB luminosities. The
  distribution of these luminosities can independently  be used to
  get information about the luminosity function.

In this paper we constrain the luminosity function of GRBs using two 
approaches: (a) using the redshift data for a sample of 16 GRBs from
afterglow measurements; (b) using the number-count v/s flux 
(i.e,  ${\cal N} \hbox{--}  F$ ) relation for GRBs
in the current Burst and Transient Source Experiment (BATSE) catalog. 
In doing so, we consider various
luminosity functions  (viz. Schechter, scale-free and log-normal) 
each with
`no-evolution' and some of the evolutionary models consistent 
with star-formation history of the universe.
Our aim is to identify the class of luminosity functions which are
consistent with both the ${\cal N}\hbox{--}  F$
relation for BATSE GRBs and the luminosity distribution of  
the  sample of GRBs
whose redshifts are available from the afterglow observations.

In \S 2 we 
briefly review the ${\cal N} \hbox{--}  F$  method and describe models
of luminosity function and GRB number density evolution.
In \S 3 we discuss the afterglow redshift data and
the Likelihood method used for 
extracting information about the luminosity function parameters
from this data. The results are presented 
and discussed in \S 4. Throughout this paper
we assume a cosmological model with $\Omega_m=0.3 \>,\Omega_{\Lambda}=0.7 \>,
H_0=65 \, {\rm km \, sec^{-1} \, Mpc^{-1}}$; this model is favoured by recent observations
(Perlmutter {\it et al.} 1998, de Bernardis {\it et al.} 2000,
Freedman {\it et al.} 2000).

\section{The   \boldmath {${\cal N}({>}F)\hbox{--}F$} of BATSE GRBs}
        
For calculating the number of bursts exceeding a given
flux $F$, ${\cal N}({>}F)$,   from BATSE sources 
 we first need to 
define `flux' of a GRB. We use the BATSE peak photon flux (Paczy\`nski  1995 )
averaged over the trigger time of  1.024 seconds  as being representative of GRB flux. 
 BATSE
reports  peak photon flux, $F$,   integrated  over an energy range from 
$50 \, \rm kev$ to $300 \, \rm kev$ \footnote{for more details 
see {\tt http://gammaray.msfc.nasa.gov/batse/data/}} . This is related to the photon
luminosity $L$ (in  photons  $\rm sec^{-1}$) as (see Appendix A for a derivation):
\begin{equation}
        F = {L (1+z)^{2- \alpha} \over 4 \pi D_L^2}
\label{eq:a1}
\end{equation}
Here  $F$ is in units $\rm cm^{-2} sec^{-1}$.  $\alpha$ is the 
spectral index of intrinsic photon luminosity. 
We take $\alpha = 2$  as it provides
a reasonable fit to the burst spectra (for more details and caveats 
see Band {\it et al. \/} 1993. Note that the spectral index $\alpha$
here is  the high-energy spectral index $\beta$ as defined in
Band {\it et al.} 1993).  The luminosity distance $D_L = r (1+z)$, $r$
is the coordinate distance to the  GRB at a redshift $z$. For flat cosmologies 
i.e. $\Omega_{\rm Total} =1$, it is given by:
\begin{equation}
r = H_0^{-1} \int_0^z {dz' \over (\Omega_m (1+z')^3 + \Omega_\Lambda)^{1/2}}
\label{eq:a1p}
\end{equation}
Here the Hubble length $H_0^{-1} = 9.1 \times 10^{27} h^{-1} \, \rm cm$.
 $\Omega_m$ and $\Omega_\Lambda$ are 
the present energy densities (in units of the critical density) of the 
non-relativistic matter and the cosmological constant, respectively. 
Observations of SNIa at high redshifts and anisotropies of Cosmic
Micro-wave Background Radiation (CMBR)
are consistent with a flat cosmological model with  $\Omega_\Lambda \simeq 
 0.7$   (Perlmutter {\it et al.} 1998; de Bernardis {\it et al.} 2000). 

In current BATSE catalog peak fluxes are available for 2093 events.
We   take  $F \ge 0.4 \, \rm cm^{-2} \ sec^{-1}$
for comparing  theoretical predictions with
the observed ${\cal N}({>}F)\hbox{--}F$  from the BATSE catalog
(Loredo  \& Wasserman 1998). This requirement leaves 
$\simeq 1790$ burst for our analysis. These GRBs are divided into
170 flux bins for computing the number count--flux relation.

The number count  ${\cal N}({>}F)$  can be  expressed as:
\begin{equation}
{\cal N}({>}F) = 4 \pi \int_0^{z_{max}} dr \, r^2 \int_{4 \pi D_L^2 F}^\infty dL \, \, n(L,z)  (1+z)^{-\alpha}.
\label{eq:a11} 
\end{equation} 
Here
\begin{equation}
n(L,z)= \phi_*(z) \phi(L) dL
\end{equation}
is the  comoving number density 
of GRBs in a given luminosity range $L$ to $L + dL$ at a given redshift. 
$\phi(L)$ is the luminosity function, defined such that
$\int_0^\infty \phi(L) dL = 1$.   $\phi_*(z)= \phi_*(0)\times (1+z)^\gamma$ is the total comoving number  density of GRBs at  redshift $z$. The factor of
$(1+z)^{-\alpha}$ in Eq.~(\ref{eq:a11}) gives the `k-correction'
which  corrects for the difference between the  emitted and 
the observed wavelength.

The functional form of the luminosity function of the GRBs
is unknown. In the framework of fireball models, the luminosity
function of GRBs is expected to be broad with luminosity width of
nearly 2 orders of magnitude (Kumar \& Piran 1999). 
For our study, we assume  several different  luminosity functions, $\phi(L)$:

{\it Log-Normal distribution function}: 
\begin{equation}
\phi(L) = {\exp(-\sigma^2/2) \over \sqrt{2 \pi \sigma^2} } \exp\left [ -(\ln (L/L_0))^2/(2\sigma^2) \right ] {1 \over L_0}
\label{eq:a2}
\end{equation}
$\sigma$ and $L_0$ are  the width and the average  luminosity of the
luminosity function, respectively (for previous
use of this luminosity function for GRB studies see e.g. Woods \& Loeb 1995).
The log-normal distribution  is  also 
representative of the luminosity function of  spiral galaxies (for 
details see Bingelli, Sandage \& Tamman  1988). 

{\it  Schechter distribution}: 
\begin{equation}
\phi(L) = A \left ({L \over L_*} \right )^{-\beta} \exp \left ( -L /L_* \right) {1 \over L_*} 
\label{eq:a3}
\end{equation}
$A$ is a normalizing constant. The entire  galaxy population  can be
roughly represented by this function
(for  details see Bingelli {\it et al. \/} 1988). We restrict ourselves
to $\beta \le 1$ in this paper.

{\it Scale-free luminosity function}:
\begin{equation}
\phi(L) = A \left ( {L \over L_*} \right )^{-\beta} {1 \over L_*} \quad \hbox{for }\, L_{\rm min} \le L \le L_{\rm max}
\label{eq:a4}
\end{equation}
$A$ is a normalizing constant. Several authors have used this class of
luminosity functions for GRB analyses (see e.g. Schaefer 2000).

${\cal N}({>}F)$ depends on a number of parameters: (a) cosmological 
parameters through $D_L$ and $r$.  
We fix the cosmological parameters to their most favoured values. 
(b) $z_{\rm max}$,
   the maximum redshift of GRBs, (c) $\phi_*(z)$, which gives
the redshift dependence of the GRB population, (d) the parameters 
of  the luminosity function. Our aim
is to obtain the most-favoured values of the luminosity 
function parameters.  After fixing the cosmological model and the 
GRB spectral index, the most important remaining 
uncertainty comes from the evolution of GRB population.
One obvious choice is the `no evolution' model. In this 
model, $\phi_*(z)$ is independent of $z$. However 
afterglow observations give circumstantial evidence that
GRBs might be associated with  star-forming regions
(Kulkarni {\it et al. } 2000 and references therein). Such 
an association might mean that GRB population  trace the 
star-formation history of the universe. Therefore we 
also consider several  other models, which are consistent with
the star-formation history of the universe: 

\begin{description}

\vspace{.5cm}
\item[Model I:]  No Evolution model

\vspace{.5cm}
\item[Model II:]    The  GRB population evolves as  the 
luminosity density at $1600 \, {\rm \AA}$, which is a good
tracer of star-formation history 
(Madau {\it et al. \/} 1998, Madau {\it et al. \/} 1996).
In this case the comoving number density of GRBs increases by a 
factor of $\simeq 10$ from $z \simeq 0$ to $z \simeq  1.5$ 
(Lilly {\it et al. \/} 1996)  and 
then decreases approximately $\propto (1+z)^{-1}$
at higher redshifts  (Figure~3 of Madau {\it et al.} 1998).

\vspace{.5cm}
\item[Model III:]  To  account for  the possible dust-extinction in 
the star-forming regions for $z \ge 1.5$ (Conolly {\it et al.} 1997), 
we assume that the star-formation rate grows
for  $z \le  1.5$ as in the previous model, but remains
constant at the value of  $z = 1.5$ till $z \simeq 3$,  and then 
  declines as $\propto  (1+z)^{-1}$ at larger 
redshifts. 

\vspace{.5cm}
\item[Model IV:]  We consider an extreme  model that,
 given the uncertainty 
in the evolution of the star-formation rate at $z \ge 1.5$,
 is barely consistent  with the star-formation history of the 
universe. 
In this model, the comoving number 
density of GRB population evolves 
$\propto (1+z)^{3.5}$ for $z \la  1.5$ (Lilly {\it et al. \/} 1996). 
The comoving 
number density remains at the value of $z = 1.5$ for 
$z \le 5$ and then it gradually declines at $\propto (1+z)^{-0.5}$.
 In this model
much of star-formation at redshifts $\ga 1.5$ occurs in regions
highly shrouded by dust which allows a high star-formation
rate to be compatible with the star-formation rate observed in
the Hubble Deep Field UV drop-out galaxies. However, further increase in the star-formation
rate, which must be accompanied by a suitable increase in the dust 
content, might be incompatible with the observation of Far Infra-red
background (Puget {\it et al.} 1996, Guiderdoni {\it et al. 1997}).
\end{description}

\section{Observed GRB redshifts}

In Table~1  we list the redshifts of GRBs used to calculate the
luminosity distribution of GRBs. The GRB at $z = 0.008$ (GRB 980425) has
been excluded from this list, as it is probably associated with a supernova
(Galama {\it et al.} 1998) and therefore corresponds to a different
population of GRBs (Kulkarni {\it et al. \/} 2000).
Three GRBs whose   redshifts are not certain (GRB 980326: $z \simeq 1$;
GRB 980329: $z \le 3.5$; GRB 990507: $z \simeq 0.25$) are also  excluded.

 Gamma-ray fluxes ($\rm photons \, \, cm^{-2} \, sec^{-1}$)
corresponding to these 16 GRBs are  listed in Table~1:
Eight of these are taken from BATSE flux table (integrated over 1024 milliseconds).
The gamma-ray fluxes of remaining GRBs, triggered  
either by BeppoSAX or IPN satellites are in different energy band and therefore
had to be extrapolated to energy range of BATSE. The band-pass conversion
is achieved using photon luminosity spectral index, $\alpha = 2$ (see appendix B).
We adopt the values from Schaefer (2000) for three of the GRBs.

From the measured redshifts and observed $\gamma-$ray fluxes,
the total energies of GRBs, assuming isotropic emission,  is in the
range $10^{51-54}$ erg. The production of such large energies pose serious problems
in theoretical modelling. One of the proposal to resolve the energy crisis
was to have the emission collimated into jets, which would lower the
total energies by a factor of few hundreds (Rhoads 1999; Piran 1999).
The GRB afterglow observations have shown evidences for beaming 
in the light curves in 5 cases : 
GRB990123 (Holland 2000; Kulkarni 1999),
GRB990510 (Holland 2000, Harrison 1999),  
GRB 991216(Halpern 2000), 
GRB 000301c (Sagar {\it et al. \/} 2000, Berger {\it et al. \/} 2000 ) 
and GRB 000926 (Price {\it et al. \/} 2000).
The opening angle $\theta_\circ$ of the jet may be calculated from Eq.~(1) of
Sari {\it et al. \/} (1999) knowing the break in light curve. 
The beaming factor in Table~1.
is $\approx 2/\theta_\circ^2$ and has been used to apply the corrections
in luminosities in 5 cases.

The  probability that burst of a given flux will occur in
a redshift range  $z$ to $z + dz$ is given by: 
\begin{equation}
p(z) dz = \phi(L) \times {dL \over dz} dz
\end{equation} 
$\phi(L)$  is given by Eqs.~(\ref{eq:a2}), ~(\ref{eq:a3}), or
~(\ref{eq:a4}), and $L$ for a given 
flux $F$ and redshift is determined by Eq.~(\ref{eq:a1}). 

The likelihood that the observed GRB luminosities were 
drawn from a given luminosity function 
is:
\begin{equation}
{\cal L}(a_k) = \prod^{16}_{i= 1} p(z_i,F_i)
\end{equation}
We maximize the likelihood function with respect to the parameters
of the luminosity function, $a_k$, e.g.  $a_k =  \{L_0,\sigma \}$ 
for the Log-normal luminosity function. It should be noted
that parameters estimated using this method do not 
depend on the number density $\phi_*(z)$  of the GRBs.

\section{Results}

We show our results for the log-normal luminosity function  in Figure~1. 
 We plot the allowed region in the $L_0\hbox{--}\sigma$
plane from both the observed ${\cal N}({>}F)\hbox{--}F$  from BATSE catalog
and the afterglow redshift data. The allowed region from
requiring consistency between the observed ${\cal N}({>}F)\hbox{--}F$  relation 
 and the theoretical number counts (Eq.~\ref{eq:a11}) corresponds to the area in which the 
K-S probability  $P_{\rm ks} \ge 0.01$ (see e.g. Press {\it et al.} 1992 for 
details on the K-S test).

The best fit value of $\sigma$ and $L_0$ 
from the afterglow redshift data are: $\sigma = 1.7$ and $L_0 = 3 \times 10^{57} \,sec^{-1}$
without the beaming corrections and $\sigma = 2$ and $L_0 = 2 \times 10^{56}\,\rm sec^{-1}$
  with the beaming corrections.    We show in Figure~1
the region within which  the value of the likelihood function is
$\ge 10^{-4}$ times the value at the maximum. This correspond roughly
to  99\% confidence level for a two-parameter fit (Press {\it et al. \/} 1992).
(We do not calculate the joint confidence levels using the 
usual Fisher matrix approach because  the Likelihood function is
not a joint Gaussian distribution and also because the 
Likelihood function is very broad near the maximum). 

Results for other assumed  luminosity functions are shown in
Figure~2, 3 and 4, using the same criteria for showing
the allowed regions as given in the preceding paragraph.  
The results are shown for only two
representative scale-free models. The best-fit values of 
parameters 
from GRB redshift data are: $\beta = 0.6$, 
$L_* = 3 \times 10^{59}\,\rm sec^{-1}$ without beaming correction and 
$\beta = 0.6$, $L_* = 2 \times 10^{59}\,\rm sec^{-1}$ with beaming correction 
(Schechter luminosity function);
 $\beta = 0.65$, $L_* = 5 \times 10^{56}\,\rm sec^{-1}$ without 
beaming corrections and   $\beta = 1.05$, $L_* = 3.8 \times 10^{56}\,\rm sec^{-1}$ 
(scale-free model with
$L_{\rm min} \simeq L_*$ and $L_{\rm max} \simeq 10^3 L_*$);
$\beta = 0.65$, $L_* = 6\times 10^{57}\,\rm sec^{-1}$ without beaming 
correction and $\beta = 1.15$, $L_* = 1.2\times 10^{58}\,\rm sec^{-1}$  
 with beaming correction (scale-free model with
$L_{\rm min} \simeq 3 \times 10^{-2} L_*$ and $L_{\rm max} \simeq 10^2 L_*$).

 As seen in the Figures,
it is possible to  explain the observed ${\cal N}({>}F)$--F 
relation using any of the luminosity functions we assume.
Further the  allowed regions  
do not strongly constrain either the luminosity of the GRB
or the width of the luminosity function 
(Loredo \& Wasserman 1998, Schmidt 2000). For  any 
luminosity function the allowed range of luminosities span
atleast a decade depending on the evolution of 
GRB population (Figure~1). It can be noticed that
stronger evolution in the GRB population results in 
a higher average luminosity (Schaefer 2000). $z_{\rm max}$, the 
maximum redshift of GRBs is not an important parameter so
long as $z_{\rm max} \ga 5$. It is because even the faintest
bursts we consider in our analysis
($F \simeq 0.4 \rm \, cm^{-2} \, sec^{-1}$)
come from $z \la 5$ for much of the allowed luminosity
range.  

The GRB redshift data give  a   large range of photon 
luminosities ($10^{57} \, {\rm sec^{-1}} \la     L \la     10^{59} \, \rm sec^{-1} $ 
 in photons~$\rm sec^{-1}$). Therefore it is natural
to expect that the underlying luminosity function
is broad. It is clearly seen in the figures. It should
be noted that the GRB 
luminosities implied by the ${\cal N}({>}F)$--F relation 
are not at variance with the observed GRB luminosities. 
However,  as  seen in the figures,  these two   observations
imply two different set of luminosity  function parameters in most cases.

The log-normal is the only luminosity 
function that is compatible with both BATSE number count
and the afterglow redshift data.  This agreement is 
independent of the GRB evolution model. The agreement between the two 
data sets requires $10^{55} \, {\rm sec^{-1}} \la L_0 \la  10^{56} \, {\rm sec^{-1}}$ and $2 \la \sigma \la 3$. For other luminosity functions
the agreement between the two observations becomes better as 
 GRB evolution becomes stronger (Figures~2,~3, and ~4). However  the 
requirement that the two regions  overlap can  be fulfilled only if 
the GRBs  evolve at 
a rate much stronger than the star-formation rate of the universe. 

It is seen in Figures~1, 2, 3, and 4 that the beaming correction
does not make any quantitative  difference to our results. In almost
all the cases the 99\% region from the afterglow data becomes
smaller after the beaming correction, but it is difficult 
to draw any conclusions from it. A particularly interesting case
is one of the scale-free models (Figure~4) in which the 
beaming correction reduces the area of the 99\% region quite
dramatically. However this is owing to the fact that the beaming 
correction increases the luminosity width of the observed GRB afterglows,
and the model in question has a luminosity width of 1000 which matches
the luminosity width of the GRBs. Therefore only of a very small 
range of $L_*$ is compatible with observations.

{\bf Selection effects}: 
One possible reason for the  discrepancy/agreement  between
the number count--flux and GRB redshift data results 
could be selection effects.
Most of  the redshift determinations have been possible owing to the
detection of X-ray counterpart of the GRB by BeppoSAX. This 
results in two types of selection effects: (a) 
BeppoSAX is not sensitive to burst duration $\le 1 \, \rm sec$
(Feroci {\it et al. \/} 1999), (b) the GRBs detected by BeppoSAX 
might be  X-ray selected though  Feroci {\it et al. \/} (1999) 
emphasize it is unlikely. 

The burst-duration selection effect might mean the GRBs for which
the redshifts have been determined  belong to a different 
population of GRBs. 
The burst duration distribution for the BATSE bursts is 
observed to be bi-modal  (Fishman \& Meegan 1995),
which strongly suggest that long- and short-duration bursts 
are  different populations of GRBs. In light of this it is usual
to expect that these two might have different luminosity 
functions. To check for this effect we make a sub-catalog of 
GRB bursts with duration $\ge 2 \, \rm sec$ (measured in $T90$).
As seen in Figures~1 to~4, the range of allowed
luminosity function parameters from 
the  ${\cal N}({>}F)\hbox{--}F$ relation of this sub-catalog 
are qualitatively similar to those for the entire catalog, and our
results are not sensitive to this selection criterion.
  
Another selection effect is Malmquist bias (see e.g. Sandage 1993 
for a discussion and relevant references): If the luminosity 
function of sources is calculated from a magnitude-limited sample then
the average luminosity of the estimated luminosity function exceeds
the true mean of the underlying  population. Other moments of the 
luminosity function (variance, etc.) are also affected. For GRBs,
unlike other astronomical sources,
it is difficult to establish the criterion of source selection
.i.e. it is not completely clear that the source whose luminosities
have been established are selected randomly from a flux-limited
sample. For the purposes of this paper we have taken this assumption to be
true. To assess the effect of Malmquist bias, 
 we calculate, for the luminosity functions considered in this paper, 
 the average luminosity $\langle L
\rangle_f$ and its variance 
for a flux-limited sample (see e.g. Sandage 1993).
The flux limit of the sample is taken
to be the minimum flux for which the luminosity has been determined
($\simeq 0.5 \, \rm cm^{-2} \, sec^{-1}$; Table 1).
In addition to flux, $\langle L\rangle_f$ has dependence on 
the parameters of the luminosity function  and evolutionary properties
of the whole population. We find that, within the range of relevant
luminosity function parameters and the evolution of GRB population:
\begin{itemize}
  \item[1.] The average luminosity of the flux-limited sample exceed
    the true mean by a factor of two to three.
    \item[2.] The variance of the flux-limited sample is within
      $40\hbox{--}50\%$ of the true variance.
    \end{itemize}
    This is well within the range of $99\%$ confidence contours
    from the afterglow observations (Figures~1 to~4). 
    It should be pointed out that to get complete information about the
    luminosity function, we need to calculate all the higher moments.
    However, we note that uncertainties in the first few  moments are
    far below the range of luminosity function parameters allowed by
    the likelihood analysis of the afterglow data (Figures~1 to~4). 
    This suggests that errors
    in the luminosity function parameter are more due to the smallness
    of the sample than the Malmquist bias, and our conclusion are not
    affected by this bias.

To sum up:  of the four luminosity functions considered here,
log-normal is the only luminosity function that is consistent with 
both the BATSE number count--flux relation and afterglow redshift 
data, independent of the evolution of the GRB population.  Notwithstanding 
the selection effects, the other luminosity function can be consistent 
with both observations only if the GRBs evolution far exceeds
the evolution of star-formation rate in the universe.

\section*{Appendix A}

Here we give a brief derivation of Eq.~(\ref{eq:a1}). 
Assume that a source at a redshift $z$ is emitting with  photon luminosity $L_\nu$ ($\rm photons \, \,  sec^{-1} \, Hz^{-1}$), with intrinsic spectrum:
\begin{equation} 
L_\nu = L_f \left ( { \nu \over \nu_f} \right )^{-\alpha}
\end{equation}
Here $\nu_f$ is some fiducial frequency.  
The received photon flux ($\rm photons \, \, cm^{-2} \,sec^{-1} \, Hz^{-1} $) is:
\begin{equation}
F_{\nu_0} = {L_\nu \over 4 \pi r^2}
\label{eq:app2}
\end{equation}
Here $r$, defined in Eq.~(\ref{eq:a1p}), is the coordinate  distance to 
the source from the observer at the present epoch and $\nu_0$ is  the observed frequency.  Note that there are
no factors of (1+z) in the denominator of Eq.~(\ref{eq:app2}). Integrating 
the flux over the observed band-pass   ($50 \, \rm kev$ to $300 \, \rm kev$ for
BATSE), and using $\nu/\nu_0 = (1+z)$,  we get:
\begin{equation}
F  \equiv \int F_{\nu_0} d\nu_0 = {L_f (1+z)^{-\alpha}\over 4 \pi r^2} \int \left ({\nu_0 \over \nu_f} \right )^{-\alpha} d \nu_0
\end{equation}
Defining 
\begin{equation}
L = L_f \int \left ({\nu_0 \over \nu_f} \right )^{-\alpha} d \nu_0,
\end{equation}
and using $D_L = (1+z) r$, we obtain Eq.~(\ref{eq:a1}). 

\section*{Appendix B }

Throughout this paper we take the `flux' to mean the photon 
flux averaged over 1.024~sec BATSE trigger, in the 
energy range $50\hbox{--}300 \, \rm kev$.  Out of 16 afterglows
whose redshifts have been determined, only 8 have known  BATSE fluxes.
In the other cases, the gamma-ray fluxes  are observed in other
frequency bands by either BeppoSAX or IPN satellites. To be 
consistent with our definition of flux, we extrapolate the observed 
fluxes from the observed band to the BATSE band. We briefly describe the 
method of this extrapolation in this appendix. It should be 
pointed out that the difference of trigger time between BATSE 
and other instruments must also be taken into account for extrapolating 
fluxes. However, we apply only band-pass corrections. 

Assume that a source at redshift $z$ is emitting with 
a luminosity $L_\nu$ ($\rm erg \, sec^{-1} \, cm^{-2}$)  and spectrum:
\begin{equation}
L_\nu = L_f \left ( {\nu \over \nu_f} \right )^{-\beta}
\end{equation}
$\nu_f$ is some fiducial frequency. 
The observed flux at the frequency $\nu_0 = \nu /(1+z)$ is:
\begin{equation}
F_{\nu_0} = {L_\nu \over 4 \pi r^2 (1+z)}
\label{eq:appb2}
\end{equation}
The flux integrated over a band-pass between frequencies $\nu_1$ and $\nu_2$
is given by:
\begin{equation}
F(\nu_1,\nu_2) \equiv \int_{\nu_1}^{\nu_2} F_{\nu_0} d\nu_0 = {F(z) \over \nu_f^{-\beta}} \int_{\nu_1}^{\nu_2}   \left ( {\nu_0 \over \nu_f} \right )^{-\beta} d\nu_0,
\end{equation}
with the definition $F(z) = L_f (1+z)^{-\beta -1} /(4 \pi r^2)$. We assume
$\beta =1$ in this paper (the spectral index $\alpha$ for the photon 
luminosity is related to $\beta$ as $\alpha = \beta +1$). This gives:
\begin{equation}
F(\nu_1,\nu_2) = {F(z) \over \nu_f^{-1}} \log(\nu_2/\nu_1).
\label{eq:appbma}
\end{equation}
Eq.~(\ref{eq:appbma}) can be used to convert flux observed in any band-pass
to the BATSE band-pass.

The photon luminosity ($L_\nu^p$) is related to the 
energy luminosity as: $L_\nu^p = L_\nu/(h \nu)$, $h$ being the 
Planck's constant. This relation and the methods described above as well as in
the previous appendix can be used to get the conversion between
the band-pass integrated photon flux ($F$) and the energy flux 
($F(\nu_1,\nu_2)$). The relation is (for $\beta = 1$):
\begin{equation}
F = F(\nu_1,\nu_2) {\nu_1^{-1} - \nu_2^{-2} \over h \,  \log(\nu_2/\nu_1)}.
\end{equation}

\begin{acknowledgements} 
We thank  J Greiner, R. Cowsik, and D. Hartmann 
for reviewing the manuscript and giving useful comments; A. Loeb for
suggesting us to make beaming corrections; D. Bhattacharya for useful
discussions and M. R. Kippen for information on our queries about BATSE
fluxes. We also thank the  referee for helpful comments.  
\end{acknowledgements}

\newpage

\begin{table*}
\begin{minipage}{180mm}
\vspace{3cm}
\caption[]{Redshifts and luminosities of  Afterglow GRBs }
           \label{tbl:aglow}
\vspace{.5cm}
\begin{tabular}{lllllll}
\hline
GRB & BATSE  & Redshift  & Photon flux  & Luminosity & Beaming & Ref. for\\ 
    & tr\#   &     z     & $\rm ph/s/cm^2$ & $\rm ph/s$ &factor\footnotemark{*} & 
photon flux\footnotemark{**} \\
\hline
GRB970228  &  --   & 0.695    &  10.0     & $2.34\times 10^{58}$ &  -      & 1 \\
GRB970508  & 6225  & 0.835    &   0.969   & $3.56\times 10^{57}$ &  -      & 2 \\
GRB970828  & 6350  & 0.958    &   1.5     & $7.74\times 10^{57}$ &  -      & 1 \\
GRB971214  & 6533  & 3.418    &   1.955   & $2.27\times 10^{59}$ &  -      & 2 \\
GRB980613  & --    & 1.096    &   0.5     & $3.60\times 10^{57}$ &  -      & 1 \\
GRB980703  & 6891  & 0.966    &   2.398   & $1.26\times 10^{58}$ &  -      & 2 \\
GRB990123  & 7343  & 1.6      &  16.41    & $3.0\times 10^{59} $ &  300    & 2 \\
GRB990506  & 7549  & 1.31     &  18.56    & $2.08\times 10^{59}$ &  -      & 2 \\
GRB990510  & 7560  & 1.619    &   8.16    & $1.54\times 10^{59}$ &  300    & 2 \\
GRB990712  & 7647  & 0.434    &  11.64    & $8.56\times 10^{57}$ &  -      & 2 \\
GRB991208  & --    & 0.706    &  11.2     & $2.72\times 10^{58}$ &  -      & 3 \\
GRB991216  & 7906  & 1.02     &  67.5     & $4.06\times 10^{59}$ &  200    & 2 \\  
GRB000131  & -     & 4.5      &   1.5     & $3.35\times 10^{59}$ &  -      & 3 \\
GRB000301c & -     & 2.03     &   1.32    & $4.34\times 10^{58}$ &   90    & 3 \\
GRB000418  & -     & 1.118    &   3.3     & $2.5\times 10^{58} $ &  -      & 3 \\
GRB000926  & -     & 2.066    &  10.0     & $3.45\times 10^{59}$ &  120    & 3 \\  
\hline
\end{tabular}
\vspace{.5cm}
\end{minipage}
\footnotemark{*}{References for beaming factor follow:
GRB 990123: Holland {\it et al. \/}2000; \\ 
GRB 990510: Holland {\it et al. \/}2000, Harrison {\it et al. \/}1999; \\  
GRB 991216: Halpern {\it et al. \/}2000; \\
GRB 000301c: calculated for $t_b$=4.1 days from Bhargavi \& Cowsik (2000) \\
using Eq~1. of Sari {\it et al. \/}1999;  
GRB 000926: Price {\it et al. \/}2000}

\vspace{.5cm}
\footnotetext{**}{The references for fluxes are as follows: 1. Schaefer(2000); 2. BATSE catalog; \\
3. conversion made by applying band-pass corrections as 
explained in {\bf Appendix B.} } 
\end{table*}
\vfill

\newpage

 \begin{figure*}
\centering
\includegraphics[angle=270,width=17cm]{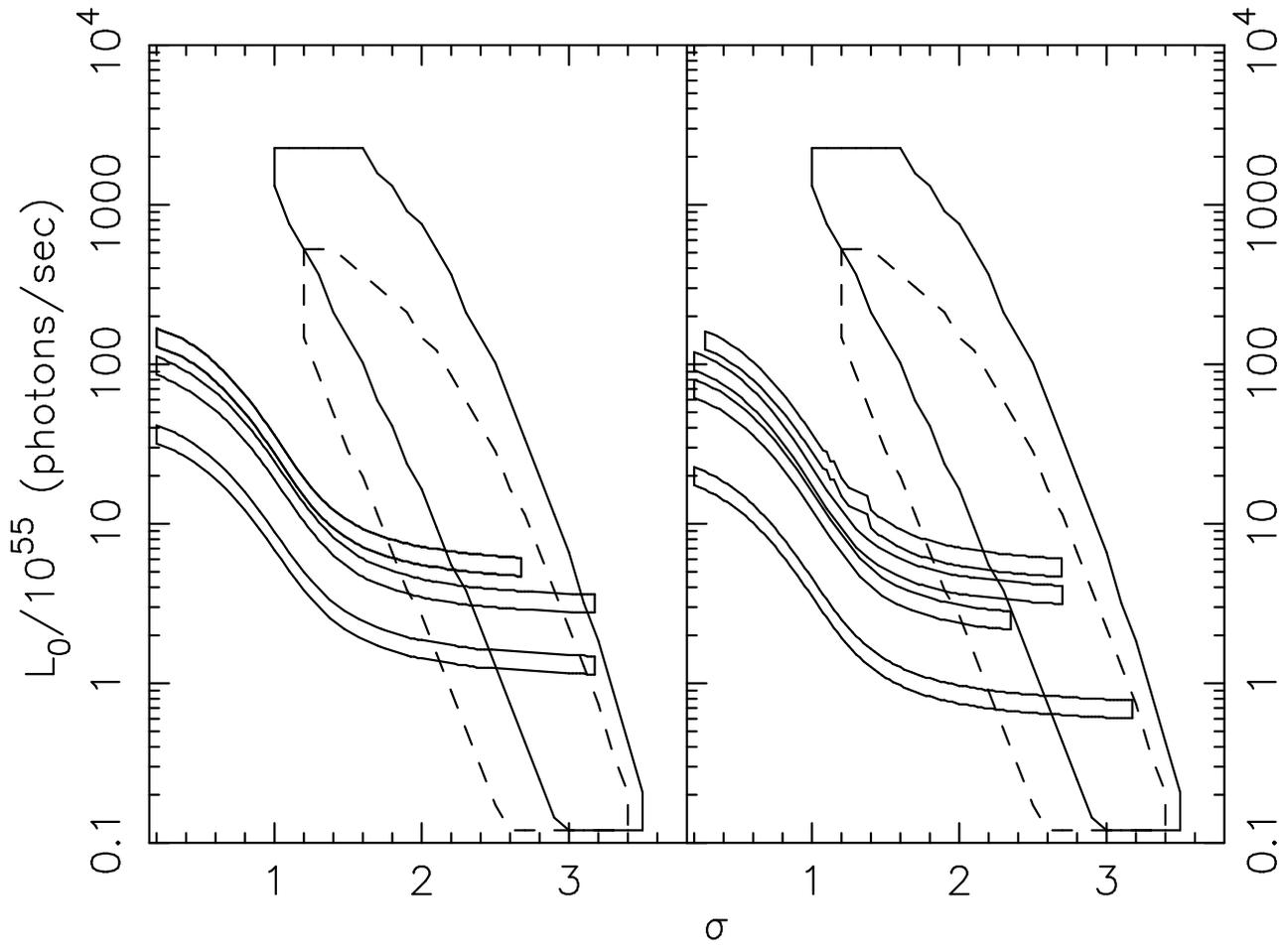}
\caption {
The results for the Log-normal luminosity function are shown.
{\it Right Panel}:
Results are shown for the full GRB sample.
The bigger regions in the center enclose the allowed region from
the afterglow observations. The region with dotted lines
corresponds to a run where beaming corrections are applied.
The 4 contours on the left side represent the region of K-S
probability $P_{ks} > 0.01$ for the consistency between observed
and theoretical number count-flux relation. They correspond, 
from bottom to top (with increasing luminosity) to four models
1-4 respectively of GRB evolution described in the text. {\it Left
  Panel}: Results are shown for the GRB sample with GRB duration (in
T90) exceeding 2~sec. Model~4 of GRB evolution is not shown.
  }
 \label{fig:f1}
\end{figure*}
 
\newpage

 \begin{figure*}
\includegraphics[angle=270,width=17cm]{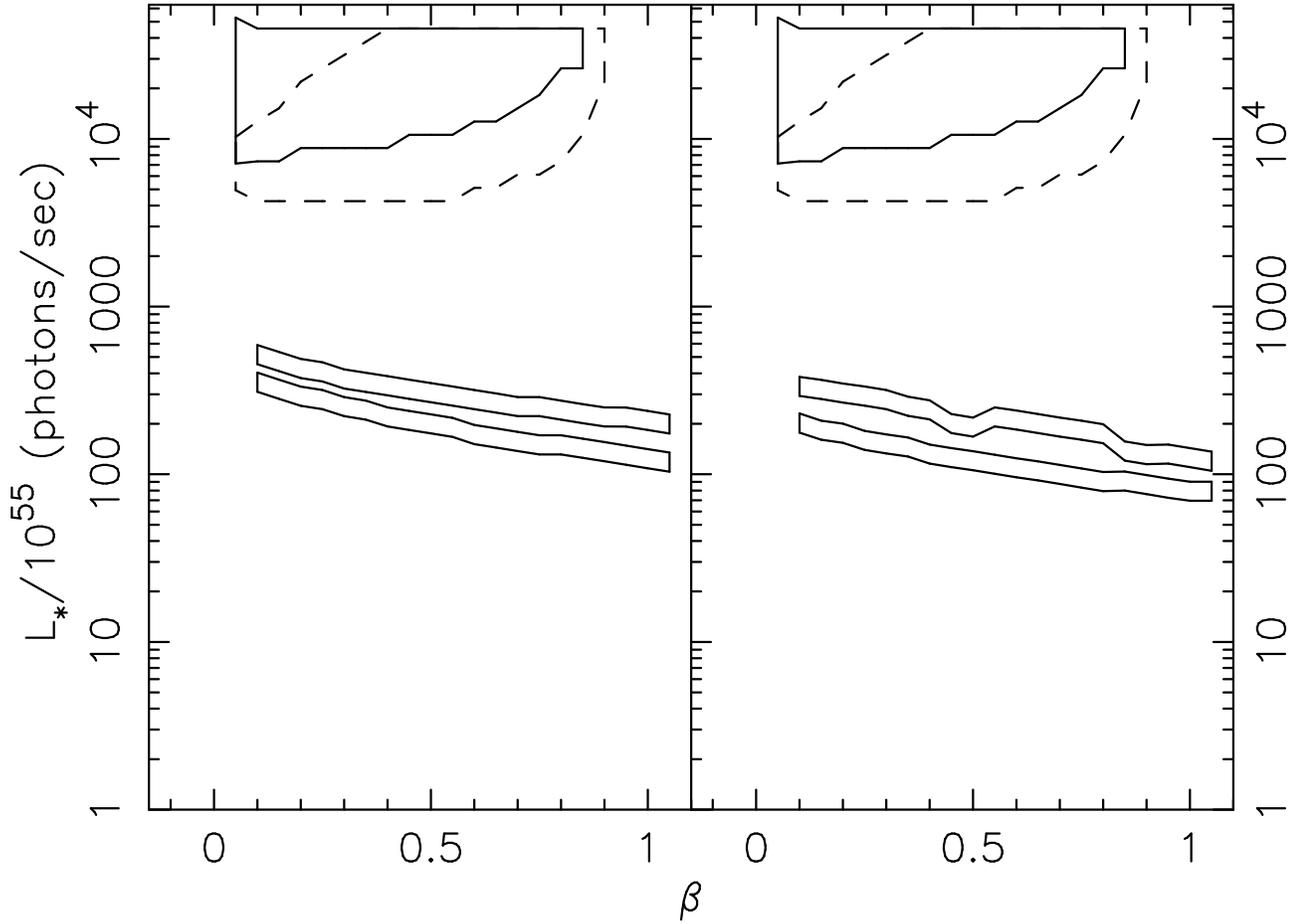}     
\caption {
The results for the Schechter luminosity function
are shown. The right  and left  Panels correspond to
the full sample and the long duration sample ($\rm T90 \ge 2\, \rm
sec$), respectively. The bigger regions at  the top of the  figure come
from the afterglow observations. The  region inside the solid (dashed)
curves correspond  to no (with) beaming correction.
 The smaller regions at the center
are    from the K-S
test for the number count--flux relation. These correspond,
 with increasing photon luminosity, to
GRBs evolution models  III and IV  discussed in the text. The
allowed regions for models I and II fall below the allowed regions
for the models shown.  }
\label{fig:f2}
\end{figure*}
 
\newpage

\begin{figure*}
\includegraphics[angle=270,width=17cm]{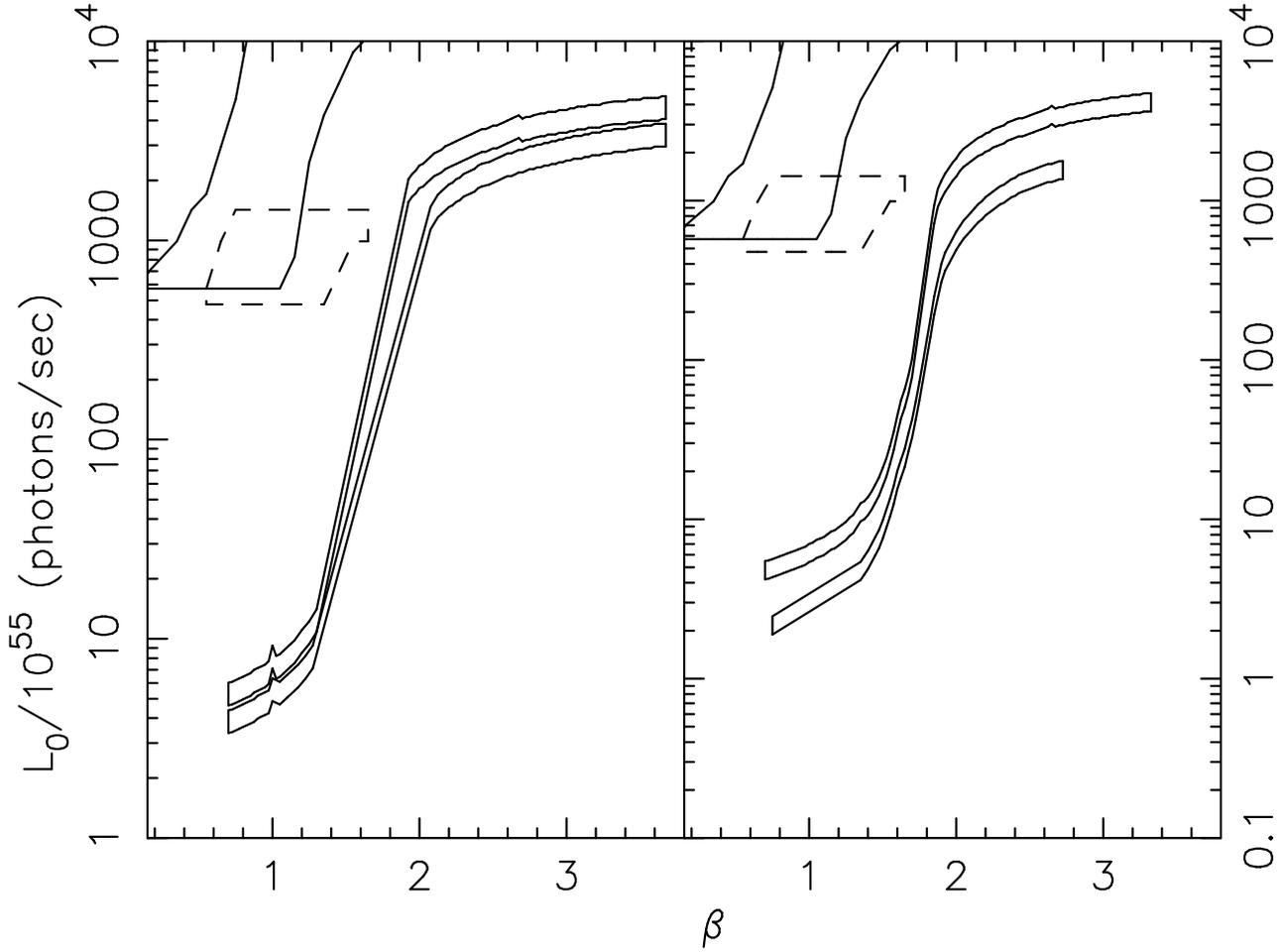}
\caption {The results for the scale-free luminosity function
are shown. In this model $L_{\rm min} = 3 \times 10^{-2} L_*$ and
$L_{\rm max} = 100 L_*$. The right  and left  Panels correspond to
the full sample and the long duration sample ($\rm T90 \ge 2\, \rm
sec$), respectively. The  regions  at  the left of the  figure come
from afterglow observations; the region  inside the solid (dashed)
curves correspond  to no (with) beaming correction.  The thin  regions
are    from the K-S
test for the number count--flux relation. These correspond,
with increasing photon luminosity, to
GRBs evolution models  III and IV  discussed in the text. The
allowed regions for models I and II fall below the allowed regions
for the models shown.  }
\label{fig:f3}
\end{figure*}
 
\newpage
 
\begin{figure*}
\centering
\includegraphics[angle=270,width=17cm]{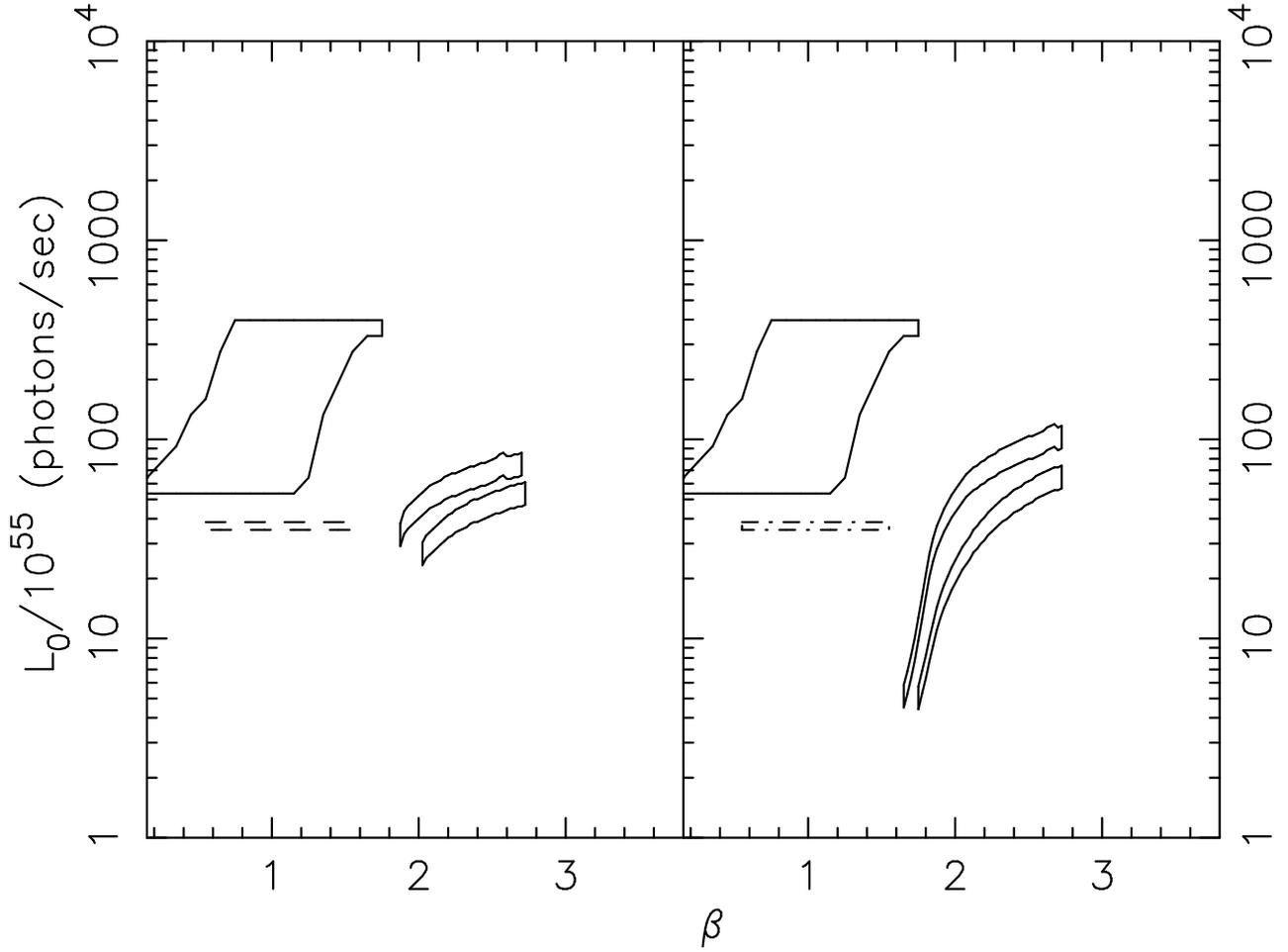}
\caption {The results for the scale-free luminosity function
are shown. In this model $L_{\rm min} =  L_*$ and
$L_{\rm max} = 1000 L_*$. The right  and left  Panels correspond to
the full sample and the long duration sample ($\rm T90 \ge 2\, \rm
sec$), respectively. The  regions at  the left of the  figure come
from afterglow observations with no (with) beaming correction
corresponding to solid (dot-dashed) curves.  Thin  regions  in the
center of the figure
are   from the K-S
test for the number count--flux relation. These correspond,
 with increasing photon luminosity, to
GRBs evolution model III and IV discussed in the text. The
allowed regions for models I and II fall below the allowed regions
for the models shown.  }
  \label{fig:f4}
\end{figure*}


\begin{thebibliography}{}
\bibitem{1}Band, D. {\it et al. \/} 1993, ApJ, 413, 281
\bibitem{} Band, D., Hartman, D., \&  Schaefer, B. 1999, ApJ, 514, 862
\bibitem{} Berger, E. {\it et al. \/} 2000, astro-ph/0005465
\bibitem{} Bhargavi, S. G. \& Cowsik, R., 2000, ApJ 545, L77
\bibitem{} Bingelli, B., Sandage,A., \& Tamman, G. A. 1988, ARA \& A, 26, 509
\bibitem{}Connolly, A. J., Szalay, A. S., Dickinson, M., SubbaRao, M. U. \& Brunner, R. J. 1997, ApJ, 486, L11
\bibitem{} Cohen, E. \& Piran, T., 1995, ApJ 444, L25
\bibitem{} de Bernardis, P. {\it et al. \/} 2000, Nature, 404, 995
\bibitem{} Feroci, M. {\it et al. \/} 1999, A\&AS, 138, 407
\bibitem{} Fishman, G. J. \& Meegan, C. A. 1995, ARA \& A, 33, 415
 \bibitem{} Freedman, W. {\it et al.} 2000, astro-ph/0012376 
\bibitem{} Galama, T. J., {\it et al. \/} 1998, Nature, 395, 670
\bibitem{} Greiner, J. 2000, {\tt http://www.aip.de/People/JGreiner}
\bibitem{} Guiderdoni, B. {\it et al. \/} 1997, Nature, 390, 257
\bibitem{} Halpern, J. P., {\it et al. \/} 2000, astro-ph/0006206
\bibitem{}  Hakkila, J. {\it et al.}, 1996, ApJ 462, 125 
\bibitem{} Harrison, F. A., {\it et al. \/} 1999, ApJ 523, L121
\bibitem{} Holland, S., {\it et al. \/} 2000, astro-ph/0010196
\bibitem{}   Horack, J. M. \& Hakkila, J. 1997, ApJ, 479, 371
\bibitem{} Kulkarni, S. R. {\it et al.} 2000, astro-ph/0002168
\bibitem{} Kulkarni, S. R. {\it et al.} 1999, Nature, 398, 389
\bibitem{} Kumar, P. \& Piran, T. 1999, astro-ph/9909014
\bibitem{} Lilly, S., Le F\`evre, O. Hammer, F, Crampton, D. 1996, ApJ,
460, L1
\bibitem{} Loredo, T. J. \& Wasserman, I. 1998, ApJ 502, 75
\bibitem{} Lubin, L. M. \& Wijers, R. A. M. J., 1993, ApJ 418, L9
\bibitem{} Madau, P., Pozzetti, L., \& Dickinson, M. 1998, ApJ, 498, 106
\bibitem{} Madau, P., Ferguson, H. C., Dickinson, M., Giavalisco, M.,
 Steidel, C., \& Fruchter, A. 1996, MNRAS, 283, 1388
\bibitem{} Mao, S. \& Paczy\'nski, B., 1992, ApJ 388, L45
\bibitem{} Paczy\`nski, B., 1995, PASP, 107, 1167
\bibitem{} Perlmutter, S. {\it et al. \/} 1999, ApJ, 517, 565
\bibitem{} Piran, T., 1992, ApJ 389, L45
\bibitem{} Piran, T. 1999, Phys. Rep., 314, 575
\bibitem{} Press, W. H., Teukolsky, S., Vellering, W. T., \& Flannery, B. P. 1992, {\it Numerical Recipes in Fortran}, Cambridge university Press 
\bibitem{} Price, P. A. {\it et al. \/} 2000, astro-ph/0012303
\bibitem{} Puget, J-. L. {\it et al. \/} 1996, A\&A, 308, L5  
\bibitem{} Rhoads, J. E. 1999, ApJ, 525, 737
\bibitem{} Sagar, R. {\it et al. \/} 2000, BASI, 28, 499
\bibitem{} Sandage, A. 1993, in Saas-Fee advance course 23,
  Eds. B. Bingelli \& R. Buser 
\bibitem{} Sari, R. {\it et al. \/} 1999, ApJ, 519, L17
\bibitem{} Schmidt, M. 2000, astro-ph/0001121
\bibitem{} Schaefer, B. 2000, ApJ, 532, L21
\bibitem{} Ulmer, A. \& Wijers, R. A. M. J., 1995, ApJ 439, 303
\bibitem{} Ulmer, A., Wijers, R. A. M. J. \&  Fenimore, E. E., 1995, ApJ 440, L9
\bibitem{} Woods, E. \& Loeb, A. 1995, ApJ, 453, 583 
\end{thebibliography}
\end{document}